\documentstyle[12pt,epsfig]{article}
\input psfig.sty
\makeatletter
\def\appendix{\par
  \setcounter{section}{0}
  \setcounter{subsection}{0}
  \def\thesection{Appendix \Alph{section}}
  \def\theequation{\Alph{section}\arabic{equation}}
  \@addtoreset{equation}{section}}
\makeatother

\def\wl{W_L^{}}
\def\zl{Z_L^{}}

\def\ttb{t\bar t}
\def\tbb{t\bar b}
\def\gev{\rm GeV}
\def\fbi{\rm fb^{-1}}

\def\mh{M_H^{}}

\def\mrho{M_\rho^{}}
\def\lsim{\mathrel{\raise.3ex\hbox{$<$\kern-.75em\lower1ex\hbox{$\sim$}}}}
\def\gsim{\mathrel{\raise.3ex\hbox{$>$\kern-.75em\lower1ex\hbox{$\sim$}}}}
\newcommand{ \slashchar }[1]{\setbox0=\hbox{$#1$}   
   \dimen0=\wd0                                     
   \setbox1=\hbox{/} \dimen1=\wd1                   
   \ifdim\dimen0>\dimen1                            
      \rlap{\hbox to \dimen0{\hfil/\hfil}}          
      #1                                            
   \else                                            
      \rlap{\hbox to \dimen1{\hfil$#1$\hfil}}       
      /                                             
   \fi}                                             %

\def\etal{{\it et al.}}

\def\tev{\,{\rm TeV}}
\def\gev{\,{\rm GeV}}

\def\to{\rightarrow}

\def\be{\begin{equation}}
\def\ee{\end{equation}}
\def\bea{\begin{eqnarray}}
\def\eea{\end{eqnarray}}
\def\atversim#1#2{\lower0.7ex\vbox{\baselineskip\zatskip\lineskip\zatskip
  \lineskiplimit 0pt\ialign{$\matth#1\hfil##\hfil$\crcr#2\crcr\sim\crcr}}}

\renewcommand{\thefootnote}{\fnsymbol{footnote}}

\begin{document} 

\begin{titlepage} 
\rightline{\vbox{\halign{&#\hfil\cr
&MADPH-00-1181\cr
&FERMILAB-Pub-00/113-T\cr
&hep-ph/0005306\cr
&May, 2000\cr}}}
\begin{center} 
 
{\Large\bf Top-quark couplings to TeV resonances\\ at future lepton colliders}

\medskip

\normalsize 
{\large T. Han$^a$, Y.J. Kim$^a$, A. Likhoded$^b$, 
and G. Valencia$^{c,d}$} \\
\vskip .3cm
$^a$Department of Physics, University of Wisconsin, Madison, WI 53706\\
$^b$Institute for High Energy Physics, Protvino, Russia\\
$^c$Department of Physics and Astronomy, Iowa State University,
Ames, IA 50011\\
$^d$Fermi National Accelerator Laboratory, Batavia, IL 60510\\
\vskip .3cm

\end{center} 

\begin{abstract} 

We study the processes 
$\wl\wl \to \ttb$  and $\wl\zl \to t\bar b\ (\bar t b)$ 
at future lepton colliders as probes of the 
couplings of the top quark to resonances 
at the TeV scale. We consider the cases in 
which the dominant low energy feature of a strongly 
interacting electroweak symmetry breaking sector 
is either a scalar or a vector resonance with mass near 1~TeV. 
We find that future lepton colliders with high energy and
high luminosity have great potential to sensitively probe 
these physics scenarios. In particular, at a 1.5 TeV linear
collider with an integrated luminosity of 200 fb$^{-1}$,
we expect about 120 events for either a scalar
or a vector to decay to $t\bar t,\ tb$.
Their leading partial decay widths, which 
characterize the coupling strengths, can be statistically 
determined to about $10\%$ level.

\end{abstract} 

\renewcommand{\thefootnote}{\arabic{footnote}} 
\end{titlepage} 


\section{Introduction}  

The mass generation mechanisms for electroweak gauge
bosons and for fermions are among the most prominent
mysteries in contemporary high energy physics. 
In the standard model (SM) and its supersymmetric 
extensions, elementary scalar doublets of the
SU(2)$_L$ interactions are responsible for the mass
generation. Yet there is no explanation for the 
scalar-fermion Yukawa couplings. On the other hand,
if there is no light Higgs boson found in the next 
generation of collider experiments, then the interactions
among the longitudinal vector bosons 
would become strong at a scale of ${\cal O}$(1 TeV)
and new dynamics must set in \cite{sews}. The fact that the top-quark
mass is very close to the electroweak scale 
($m_t \approx v/\sqrt 2$) is rather suggestive: there may be a 
common origin for electroweak symmetry
breaking and top-quark mass generation.
Much theoretical work has been 
carried out recently in connection to the top quark and
the electroweak sector \cite{topcolor,topss,liz}.

Due to the Goldstone boson equivalence theorem (ET) \cite{eqt},
the longitudinal gauge bosons ($W_L^\pm,\zl$) 
resemble the Goldstone
modes ($w^\pm,z$) at energies much larger than their mass $M_W$
and thus faithfully reflect the nature of the electroweak 
symmetry breaking (EWSB). To study the EWSB sector in
connection with the top quark, a sensitive probe is to 
produce a top quark via longitudinal gauge boson 
scattering \cite{cfh}
\begin{eqnarray}
\label{wwtt}
&&W_L^+ W_L^-,\ \zl\zl \to t\bar t,\\
&&W_L^\pm \zl \to t\bar b, \bar t b.
\label{wztb}
\end{eqnarray}
These processes will receive significant enhancement if
there are underlying resonances in this sector that couple 
to both Goldstone bosons and the top quark. In particular, it is 
interesting to note that both a scalar (Higgs-like) resonance 
and a vector (techni-rho-like $\rho^0$) resonance would 
contribute to $W_L^+ W_L^- \to t\bar t$ in process (\ref{wwtt}); 
while only a vector (techni-rho-like $\rho^\pm$) resonance 
would significantly enhance process (\ref{wztb}).

These processes can be effectively realized at high energy 
lepton colliders via nearly collinear gauge boson radiation
\begin{eqnarray}
\label{eenntt}
e^+e^- \to\nu \bar\nu\ W_L^* W_L^* \to \nu \bar\nu\ t\bar t,\\
e^+e^- \to e \nu\ W_L^* Z_L^* \to e \nu\ t b,
\label{eeentb}
\end{eqnarray}
where $e \nu\ t b$ generically denotes $e^- {\bar\nu}\ t{\bar b}$
and $e^+ \nu\ {\bar t}b$.
Within the effective $W$-boson approximation (EWA) \cite{ewa}, the 
$t\bar t$ production of Eq.~(\ref{wwtt}) was first calculated 
in Ref.~\cite{russel} and then in Ref.~\cite{steveg}. In the approach
of an electroweak effective Lagrangian, they were studied
in Ref.~\cite{eff}. A full evaluation of the SM diagrams was 
performed for $t\bar t$ production
at a 1.5 TeV linear collider \cite{timb,cp2} 
and at a multi-TeV muon collider \cite{muon}.
Effects on $W_L^+ W_L^- \to t\bar t$ from other strongly
interacting dynamics were recently discussed in \cite{mikep,peskin}.

In this paper, we carry out a comprehensive evaluation for 
processes (\ref{eenntt}) and (\ref{eeentb}) within and beyond the SM. 
In Sec.~2, we first formulate the effective interactions for a 
strongly-interacting electroweak sector (SEWS)
including the heavy top quark (Top-SEWS). 
We parameterize the sector
in a (relatively) model-independent way by introducing
a heavy scalar or a heavy vector to unitarize (up to a few TeV) 
the universal low-energy amplitudes \cite{let}. We also 
comment on the current low-energy constraints on this
sector. In Sec.~3, we perform detailed numerical analyses
for the Top-SEWS signal and the SM backgrounds. We present
our results at an $e^+e^-$ linear collider with a 
center-of-mass energy $\sqrt s=1.5$ and also illustrate
some results at a lepton collider with CM energy of 4 TeV. 
We find that the future high energy lepton colliders
have substantial potential to explore the Top-SEWS sector
to great precision. Section 4 contains our conclusions.
Some useful formulae are presented in \ref{app}.

\section{Effective Interactions in the Top-SEWS Sector
\label{sews-top}}

\subsection{Low Energy Amplitudes}

The low energy behavior of the scattering amplitudes 
for the processes $V_L V_L \rightarrow q\bar{q}$ is 
determined by symmetry, and is 
the same in all models in which the electroweak symmetry 
is broken by a strong interaction. If we parameterize the 
scale of this strong interaction with the mass of its 
lightest resonance $M_R$, then in the region where 
$m_W \ll \sqrt{s} \ll M_R$, the amplitudes are dominated 
by the low energy theorems (LET). These low energy theorems 
can be obtained in a simple manner with the use of the 
Goldstone-boson equivalence theorem as in Refs.~\cite{eqt,cfh}. 

The framework for a model-independent analysis is 
that of effective Lagrangians. Within this framework, the low
energy amplitudes arise from the lowest dimension operator 
one can construct that respects the symmetries of the standard model. 
The non-renormalizable effective Lagrangian responsible for the 
low energy interactions of the would-be Goldstone bosons $w^\pm,z$ 
is given by
\begin{equation}
{\cal L} = {v^2\over 4}{\rm Tr}\biggl(\partial_\mu U^\dagger 
\partial^\mu U\biggr) 
\label{letel}
\end{equation}
where $U=\exp(iw^i\tau_i/v)$ \cite{let,order2}, 
with ${\rm Tr}\tau_i\tau_j = 2\delta_{ij}$  
and $v\approx 246$ GeV. The minimal interactions 
between third generation fermions and the would-be Goldstone 
bosons are those of the standard model in the limit $M_H\rightarrow 
\infty$ and can be found, for example, in Refs.~\cite{cfh,xinmin}. 
We work in the limit $m_b=0$ and find (for terms with up to two 
$w^\pm,z$),
\begin{equation}
{\cal L} =
-i{m_t\over v}\ z\ \overline{t} \gamma_5t +
{i m_t\over\sqrt{2}v}\ \biggl(w^+ \overline{t}(1-\gamma_5)b
-w^- \overline{b}(1+\gamma_5)t
\biggr) +{m_t\over v^2}\ \overline{t}t\biggl(w^+w^-+{1\over 2}zz\biggr).
\label{cfhfl}
\end{equation}

As a benchmark we will consider the low energy amplitudes (LET) 
defined as the dominant terms in the limits 
$M_W \ll (\sqrt{s}, ~m_t) \ll (M_R, ~4\pi v)$. 
Taking the scattering angle $\theta$ to be that between 
the momentum of the $w^-$ (or $z$) and the top quark in the center 
of mass frame, we find for the neutral channels,
\begin{eqnarray}
{\cal M^{++}}_{LET}(ww\to \ttb)&=&
-{\cal M^{--}}_{LET}(ww\to \ttb) = -{m_t \sqrt s\over v^2}\beta_t
\label{letsmh}\\
{\cal M^{\pm\mp}}_{LET}(ww\to \ttb)&=& {m_t^2 \over v^2}{(1\pm\beta_t)
\sin\theta\over 1+\beta_t\cos\theta -2 m_t^2/s}
\nonumber \\
{\cal M^{++}}_{LET}(zz\to \ttb)&=&
-{\cal M^{--}}_{LET}(zz\to \ttb) = -{m_t \sqrt s\over v^2}\beta_t
\nonumber \\
{\cal M^{\pm\mp}}_{LET}(zz\to \ttb)&=& -{m_t^2\over v^2}
{\sin\theta\over 1+\beta_t\cos\theta},\nonumber
\end{eqnarray}
where $\beta_t=(1-4m_t^2/s)^{1/2}$.

For the charged $wz \to \tbb$ channel we define the angle $\theta$ between 
the momentum of the $z$ and the top quark in the center of mass
frame, to obtain
\begin{eqnarray}
{{\cal M}^{++}}_{LET}(w^+ z\to \tbb)&=&
{\sqrt 2 m_t^3\over v^2 \sqrt{s}} { \beta_m (1+\cos\theta)
\over [\beta_m^2 (1-\cos\theta)+2m_t^2/s]}  \nonumber \\
{{\cal M}^{-+}}_{LET}(w^+ z\to \tbb)&=&
-{\sqrt 2 m_t^2 \over v^2}{\beta_m  \sin\theta
\over {[\beta_m^2 (1-\cos\theta)+2m_t^2/s]}}  .
\end{eqnarray}
In writing these equations we took $M_W=m_b=0$, and
$\beta_m= (1-  m_t^2/ s)^{1/2}$. We present some 
details in \ref{app}.

The low energy amplitude for the process $ww\rightarrow t\bar{t}$ 
in Eq.~(\ref{letsmh}) grows linearly with energy and
violates partial wave unitarity at an energy $\sqrt{s_{ww}}\approx 3$~TeV 
\cite{ac,mvw}. Similarly, the low energy amplitude for the 
process $ww \rightarrow ww$ violates unitarity around 1.2~TeV 
\cite{order2,bdv}. The energy scale at which these violations of 
unitarity take place can be interpreted as the scale at which 
new physics must come into play. For our present purpose, the 
new physics that restores unitarity will be either a scalar or a 
vector resonance and we discuss these two cases in the next two 
sections. To maintain a model-independent discussion, we introduce the 
resonances through effective (non-renormalizable) low energy 
Lagrangians. Therefore, after inclusion of the resonance, the 
partial wave amplitudes will still violate unitarity at higher
energies. We will 
adjust the couplings of the resonances so that violation of unitarity 
does not occur until a scale between 2~and 5~TeV. With this 
prescription, we describe the phenomenology of a 
resonance which is the dominant dynamical feature of 
a strongly interacting electroweak symmetry breaking sector 
(SEWS) below 2~TeV.

\subsection{Scalar Resonance
\label{scalar}}

The interactions between the standard model gauge bosons and 
a generic scalar resonance $S$, have been considered in 
Ref.~\cite{baggeretal}. The leading order effective Lagrangian 
for these interactions contains two free parameters: the 
resonance mass $M_S$ and a coupling constant $g_S$ that can 
be traded for the width of the new resonance into $W$ and $Z$ 
pairs. The effective Lagrangian for the scalar resonance and 
its interactions with the $w^\pm,z$ is given by
\begin{equation}
{\cal L}= {1\over 2}\partial_\mu S\partial^\mu S-{1\over 2}
M_S^2 S^2 +{1\over 2}g_s v S{\rm Tr}\biggl(\partial_\mu U^\dagger 
\partial^\mu U\biggr) 
\label{elsr}
\end{equation}
from which one obtains that,
\begin{equation}
\Gamma_{Sww} = {3\over 32\pi}{g_S^2 M_S^3 \over v^2}.
\label{swlwl}
\end{equation}
With $g_S=1$, Eq.~(\ref{swlwl}) reproduces the width of 
the standard model Higgs boson. 

It is straightforward to compute the amplitudes for 
$ww\rightarrow ww$ scattering in this model. They are obtained 
from the Lagrangians, Eqs.~(\ref{letel}) and (\ref{elsr}). For example 
we find,
\begin{equation}
A(w^+w^- \rightarrow zz) = {s\over v^2} {s(1-g_S^2)-M_S^2\over s-M_S^2}.
\label{asfors}
\end{equation}
All the other channels can be obtained from this one by 
using custodial $SU(2)$ and crossing symmetries \cite{order2}. 
From Eq.~(\ref{asfors}), we can see that the choice $g_S=1$ 
reproduces the standard model amplitude. In this case, 
the amplitude takes a constant value at high energies as 
corresponds to the renormalizable standard model. If $g_S\neq 1$, 
however, the amplitude grows with energy violating unitarity at 
some point. To study this issue, we 
construct the $I=0, J=0$ partial wave amplitude, 
\begin{eqnarray}
a^0_0(s) &=& -{\Gamma_{Sww}\over g_S^2 M_S}
{s\over s-M_S^2+ i\Gamma_{Sww} {s\over M_S} }
\biggl(1+{s\over M_S^2}(g_S^2-1)\biggr) \nonumber \\ 
&+& {\Gamma_{Sww}\over 3 g_S^2 M_S} \biggl(
{s\over M_S^2}(g_S^2-1)-2g_S^2\biggl[1-{M_S^2\over s}\log
\biggl(1+{s\over M_S^2}\biggr)\biggr]\biggr).
\label{a0scalar}
\end{eqnarray}
To obtain this result we have introduced a scalar-resonance 
width into the $s$-channel Higgs propagator according to the 
prescription of Ref.~\cite{seymour},
\begin{equation}
{1\over s-M_S^2}\to { 1+i{\Gamma \over M_S} \over
s-M_S^2+i\Gamma {s\over M_S}}.
\label{presc}
\end{equation}
which produces a well behaved 
amplitude both near the resonance and at high energy.\footnote{
Both a constant width and an energy dependent width for a heavy 
scalar (with mass near 1~TeV) produce  undesirable features  
\cite{scottres}. We thank S. Willenbrock for pointing  
out reference \cite{seymour}.}

We first look at this result in the region above the resonance, and 
near the limit of validity of the effective theory. For 
$\sqrt{s}\gg M_S$, Eq.~(\ref{a0scalar}) reduces to
\begin{equation}
a^0_0(s) = (1-g_S^2){s\over 16 \pi v^2}-{5\over 3}{\Gamma_{Sww}
\over M_S}
\end{equation}
Once again, we see that $g_S=1$ reproduces the standard model result, 
in which the unitarity condition becomes a constraint on the 
Higgs boson mass \cite{eqt}. The first term shows how, for 
$g_S\neq 1$, the model corresponds to a non-renormalizable effective 
theory. If we demand that this term satisfy a unitarity condition
$|{\rm Re}(a^0_0(s))|< 0.5$ up to $\sqrt{s}=2\ (3)$ TeV we obtain
$0.8 \ (0.9) \leq |g_S| \leq 1.2 \ (1.1)$.
If we choose to consider all the effects of the resonance, including 
its width according to the prescription of Ref.~\cite{seymour}, we then 
find for $M_S=1$~TeV that unitarity condition
in the form $|a_0^0|<1$, is satisfied 
up to 2 (5) TeV with
\begin{equation}
0.4 \ (0.9) \leq |g_S| \leq 1.1 \ (1.02).
\label{wwgs}
\end{equation}

The effective Lagrangian describing the coupling of 
the top quark to the scalar resonance that we use is, 
\begin{equation}
{\cal L}= - \kappa {m_t\over v} S\bar t t 
\label{lstt}
\end{equation}
where $\kappa$ is a new coupling that can be traded 
for the width of the scalar into top-quark pairs, 
\begin{equation}
\Gamma_{S\ttb} = {3\kappa^2\over 8\pi}{m_t^2 M_S \over v^2}
\left(1-{4m_t^2\over M_S^2}\right)^{3\over 2}.
\label{stt}
\end{equation}
The case $\kappa=1$ corresponds to the usual standard model Higgs-boson 
coupling to the top quark. With Eqs.~(\ref{letel}), (\ref{elsr}) and 
(\ref{lstt}),  we have an effective theory involving a new generic scalar 
resonance and its couplings to would-be Goldstone bosons and 
top quarks. Using Eqs.~(\ref{elsr}) and (\ref{lstt}), and the 
Equivalence Theorem (ET), we find the matrix element for 
$\wl\wl \to \ttb$ via the scalar $S$ to be
\begin{equation}
{\cal M}_S^{++}(ww\to \ttb)=-{\cal M}_S^{--}(ww\to \ttb)= 
 g_S \kappa
{s \over s-M_S^2}{m_t\sqrt s\over v^2}\beta_t,
\label{scalarscat}
\end{equation}
here and henceforth, the double superscripts denote the
$t\bar t$ helicities.
For $\kappa = g_S = 1$, it reduces to the standard model result.
We obtain identical amplitudes for $zz\to \ttb$. 
To describe the resonance region we modify this amplitude according to 
the prescription in Ref.~\cite{seymour}.
Having chosen an $SU(2)$ singlet, electrically neutral, 
scalar resonance, there is no 
contribution to $W_L^\pm \zl \to t\bar b,\bar t b$ in this model.

There are no significant, phenomenological constraints on the 
couplings $g_S$ and $\kappa$. Following 
Refs.~\cite{cfh},~\cite{ac},~and~\cite{scott},
we can demand that the process $ww\rightarrow t\bar{t}$ satisfy 
inelastic partial-wave unitarity. By choosing the $I=0,J=0$, color 
singlet channel as in Ref.~\cite{mvw}, we find,
\begin{equation}
\biggl| 3{G_Fm_t\sqrt{s}\over 8\pi\sqrt{2}}\biggr|\biggl|
{M_S^2+s(g_S\kappa -1)\over s-M_S^2}\biggr|<0.5
\label{wwttunit}
\end{equation}
In analogy with our treatment of Eq.~(\ref{a0scalar}), we first 
concentrate in the region above the resonance. The first factor in 
Eq.~(\ref{wwttunit}) corresponds to the low energy theorem, and it 
violates unitarity at $\sqrt{s}\approx 3$~TeV. We can have the 
full amplitude satisfy unitarity up to $3$~TeV by requiring the 
second factor to satisfy
\begin{equation}
\biggl| g_S\kappa {s\over s-M_S^2} -1\biggr| <1
\end{equation}
or, for energies far above the resonance, 
$0< g_S\kappa < 2.$

If, on the other hand, we introduce the width according to the 
prescription of Ref.~\cite{seymour} and consider the full amplitude 
for $M_S=1$~TeV, unitarity is satisfied up to $\sqrt{s}=2\ (5)$~TeV if 
\begin{equation}
0\ (0.4) < g_s\kappa < 1.9 \ (1.6)
\label{gsk}
\end{equation}

Finally, we consider $\ttb \to \ttb$ scattering at energies 
above the resonance (in the same helicity, color singlet channel). 
The $J=0$ partial wave is, 
\begin{equation}
a_0(t\bar{t}\rightarrow t\bar{t}) = -{3\over 4\pi}{G_F m_t^2\over \sqrt{2}}
\kappa^2.
\label{a0tttt}
\end{equation}
If we follow Ref.~\cite{mvw} and require that 
$|{\rm Re}a_0(t\bar{t}\rightarrow t\bar{t})|< 0.5$, we obtain the 
unitarity constraint 
\begin{equation}
|\kappa|< 2.9.
\label{kappa}
\end{equation}
As we will see in our numerical studies in Fig.~\ref{sens}, this
constraint is not very restrictive.

\subsection{Vector Resonance
\label{vector}}

The interactions of standard model gauge bosons with new 
vector resonances have been described in the literature 
\cite{bagsm,bess}. In the notation of Ref.~\cite{bagsm}, two new 
parameters are introduced $a$ and $\tilde{g}$ (these 
correspond to $\alpha$ and $g^{''}/2$ from Ref.~\cite{bess}, 
respectively). The effective Lagrangian consists of a kinetic term 
for the triplet of vector resonances $\rho_\mu\equiv \rho_\mu^i\tau_i/2$ 
and an interaction term of the form
\begin{equation}
{\cal L} =-{1\over 4}av^2{\rm Tr}\biggl(
(\xi^\dagger\partial_\mu \xi +\xi\partial_\mu\xi^\dagger+
i 2\tilde{g}\rho_\mu)
(\xi^\dagger\partial^\mu \xi +\xi\partial^\mu\xi^\dagger+
i 2\tilde{g}\rho^\mu)\biggr)
\label{vecel}
\end{equation}
where $\xi=\exp(iw^i\tau_i/2v)$. 

The two new constants $a,\tilde{g}$ are related to the mass and width of the 
vector particles $\rho$ via the relations,
\begin{eqnarray}
M_\rho^2 = a v^2 \tilde{g}^2,\quad 
\Gamma(\rho^0 \rightarrow w^+ w^-) = {a^2\tilde{g}^2\over 192\pi}M_\rho.
\end{eqnarray}
In addition, when the electroweak gauge bosons are added in a 
gauge invariant manner, they mix with the new vector 
bosons \cite{bess}. In the 
charged sector, the physical states become, 
\begin{eqnarray}
V^\pm_{ph} = \rho^\pm \cos\phi + W^\pm \sin\phi,\qquad
W^\pm_{ph} = W^\pm \cos\phi - \rho^\pm \sin\phi
\label{mixing}
\end{eqnarray}
with a mixing angle given by
\begin{equation}
\tan 2\phi = {4a g\tilde{g}\over (1+a)g^2-4a\tilde{g}^2} 
\rightarrow  -{g \over \tilde{g}},
\label{mixang}
\end{equation}
the last expression being the limit as $M_\rho\rightarrow \infty$. 
Similar expressions for the mixing in the neutral sector can 
be found in the literature \cite{bess}. For our numerical studies 
we use couplings $a$, $\tilde{g}$ that have been studied in detail 
by the authors of the BESS model, and that respect partial wave 
unitarity below a few TeV. For example, 
for $M_\rho=1$~TeV and $\Gamma_\rho=30$~GeV, the partial wave 
$a_0^0$ for $ww\to ww$ scattering satisfies $|a_0^0|<1$ up to 
about 2.6~TeV, while $|a_1^1|<1$ is satisfied up to about 3~TeV 
with the simple prescription of using a constant width in the 
$s$-channel $\rho$ propagator.

We are currently interested in effective couplings 
between the new vector mesons and the top and bottom quarks. 
These couplings may arise either from the ordinary quark couplings 
to gauge bosons through the mixing of Eq.~(\ref{mixing}), or 
from novel direct couplings. The couplings induced by mixing 
are universal, common to the three generations, and can be 
bound by low energy experiments. 
More interesting is the possibility of a 
direct coupling to the third generation. 
We write these effective couplings in a generic form
\begin{equation}
{\cal L}_{eff} = 
 - \bar \psi \gamma^\mu (g_V+g_A\gamma_5)\tau^i \/\psi\ \rho^i_\mu ,
\label{ttrho}
\end{equation}
where $\bar \psi$ is the third generation quark doublet $(\bar t\ \bar b)$. 

\begin{figure}[tb]
\vskip 0.2in
\centerline{\epsfig{file=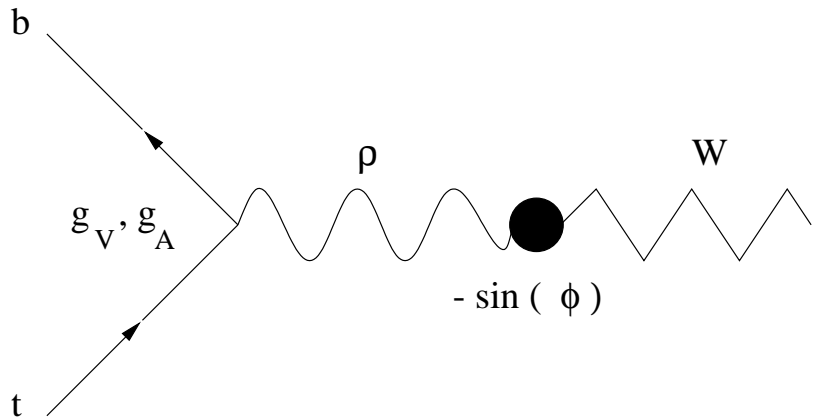}}
\caption[]{$W-V$ Mixing induced anomalous $tbW$ coupling.}
\label{mix}
\end{figure}

Unlike the case of the scalar resonance, 
there exist low energy constraints on the direct couplings of 
new vector resonances to the third generation quarks because they 
modify their couplings to the $W$ and $Z$. These deviations can be 
parameterized by the low energy effective Lagrangian
\begin{eqnarray}
{\cal L}_{eff} &=&-{g\over \sqrt{2}}\biggl[(1+\delta\kappa_L)
\bar{t}_L\gamma^\mu b_L + \delta\kappa_R\bar{t}_R\gamma^\mu b_R\biggr]
W^+_\mu +{\rm h.~c.} \nonumber \\
&& -{g\over 2\cos\theta_W}\biggl[(L_t+\delta L_t)\bar{t}_L\gamma^\mu t_L
+(R_t+\delta R_t)\bar{t}_R\gamma^\mu t_R\biggr]Z_\mu.
\label{anomcoudef}
\end{eqnarray}
For simplicity,\footnote{
The anomalous $Z$-boson couplings are obtained in a similar manner, 
but there is the additional complication of both $\rho^0$ and $Z$ 
mixing with the photon.} 
let us consider the case of anomalous $W^\pm$ 
couplings as shown in the diagram in Fig.~\ref{mix}.
We find in the limit of large vector resonance mass that 
\begin{eqnarray}
\delta\kappa_L = -{2\over \tilde{g}}(g_V -g_A), \quad
\delta\kappa_R = -{2\over \tilde{g}}(g_V +g_A).
\label{corresp}
\end{eqnarray}
There are several constraints on these couplings. The  
measurement of the rate for $b \rightarrow s \gamma$ 
restricts $|\delta\kappa_R|\lsim 0.004$ \cite{fy,lpy} but does not 
place significant constraints on $\delta\kappa_L$. An 
analysis of precision measurements at LEP \cite{dv} 
(with updated data from Ref.~\cite{pdb}) results in  
$|\delta\kappa_L| \lsim 0.1$ {\it excluding} the parameter 
$T$.\footnote{If the parameter $T$ is included, the bound could be 
stronger in certain cases. In our case, however, the induced 
anomalous couplings are of the form $\delta\kappa_L =\delta L_t$, 
and are not restricted by $T$ where they appear in the  
combination $\delta\kappa_L -\delta L_t$ \cite{dv}.} From 
future experiments, it is expected that a B factory can 
place significant further constraints of order 
$|\delta\kappa_R|\lsim 0.001$, $|\delta\kappa_L|\lsim 0.03$ \cite{hv}; 
and that single top production can yield 
$|\delta\kappa_L|\lsim 0.05$ \cite{yuan}. 
From all these numbers we conclude that a right-handed coupling 
is severely constrained and for the rest of this paper will 
concentrate on left-handed couplings exclusively. That is, we choose 
to satisfy the relation 
\begin{equation}
g_A = -g_V, 
\end{equation}
and, in keeping with the current bounds, we will concentrate 
on the case 
\begin{equation}
g_V \lsim 0.03\ \tilde{g}.  
\label{GVGA}
\end{equation}
After these bounds are imposed, partial wave unitarity does not 
place additional constraints.

When new direct couplings such as those in Eq.~(\ref{ttrho})
are introduced in a gauge invariant 
manner, as is done in the BESS model \cite{bess} (through 
their parameter $b$), 
the effective couplings $g_{V,A}$ of Eq.~(\ref{ttrho}) receive 
contributions from both the direct couplings and mixing. In this 
general case Eq.~(\ref{ttrho}) is not literally correct, as the 
charged and neutral couplings differ. We shall employ Eq.~(\ref{ttrho})
for simplicity, assuming that the direct couplings dominate. 
For example, in the heavy vector limit of the 
BESS model, they take the form
\begin{equation}
g_V = -g_A = {\tilde{g}\over 4}{b\over 1+b}.
\label{besseff}
\end{equation}
Considering both the charged and neutral vectors, this results in
\begin{equation}
\delta\kappa_L = \delta L_t = -{1\over 2}{b\over 1+b} \ \
{\rm ~and~} \ \ 
\delta\kappa_R = \delta R_t = 0.
\label{ancou}
\end{equation}
The bound $|\delta\kappa_L|\lsim 0.1$ then corresponds to 
$|b| \lsim 0.25$. In this respect, it is expected that a future 
linear collider studying the reaction $e^+e^-\rightarrow W^+W^-$ 
can probe values of $b$ some ten times smaller than this \cite{bessre}. 
Our present study is complementary in the sense that we wish to
observe the direct signal for a vector state production
at high energies.

The couplings $g_{V,A}^{}$ can be traded in favor of the physical
partial decay widths via the relations
\begin{eqnarray}
\Gamma_{\rho t\bar{t}} &=& {M_\rho  \over 4\pi}
{\left( 1- 4 {m_t^2 \over M_\rho^2} \right)}^{1 \over 2}
\left\{ {g_V^{}}^2 \left( 1+2{m_t^2 \over M_\rho^2} \right)+
{g_A^{}}^2 \left( 1+4{m_t^2 \over M_\rho^2} \right) \right\} \label{rtt} \\
\Gamma_{\rho t\bar{b}} &=& {M_\rho \over 4\pi}^{} ( {g_V^{}}^2+{g_A^{}}^2)  
{\left( 1-{m_t^2 \over M_\rho^2} \right)}^2 
\left(  2+{m_t^2 \over M_\rho^2} \right) 
\label{lastone}
\end{eqnarray}
$\Gamma_{\rho b\bar{b}}$ is the same as Eq.~(\ref{rtt}) 
replacing $m_t$ by $m_b$.
In our calculations, we consistently take $M_W=m_b=0$.

In combining the $\rho$ couplings to $\wl$ in Eq.~(\ref{vecel})
with the couplings in Eq.~(\ref{ttrho}), we obtain the scattering amplitudes
\begin{eqnarray}
{\cal M}_\rho^{++}(w^+w^-\to \ttb)&=&-{\cal M}_\rho^{--}(w^+w^-\to \ttb)
\nonumber\\
 &=& a\tilde{g}g_V 
{m_t \sqrt s\over s-M_\rho^2+iM_\rho^{}\Gamma_\rho^{}}\ 
\cos\theta\nonumber \\ 
{\cal M}_\rho^{\pm\mp}(w^+w^-\to \ttb)&=&
{a\tilde{g}\over 2} 
{s \sin\theta \over s-M_\rho^2+iM_\rho^{}\Gamma_\rho^{}}
(-g_V \mp g_A \beta_t )
\end{eqnarray}
and 
\begin{eqnarray}
{\cal M}_\rho^{\pm\pm}(w^+z \to \tbb)&=&
{a\tilde{g} \over \sqrt 2}{m_t\sqrt{s} \over 
{s-M_\rho^2+iM_\rho{}\Gamma_\rho{}}}\ \beta_m\nonumber \\
 & & \times \left\{
(\mp{g_V}+g_A)(1+\cos\theta) + (\pm{g_V} - g_A)
\right\}  \nonumber \\
{\cal M}_\rho^{\pm\mp}(w^+z \to \tbb)&=&
{a \tilde{g} \over \sqrt 2}{s \over {s - M_\rho^2 + iM_\rho{} \Gamma_\rho{}}}
\beta_m (g_V \pm g_A) \sin\theta.
\end{eqnarray}
Unlike the case of the scalar 
resonance, these amplitudes do not grow with energy. For this reason, 
partial wave unitarity does not place any significant constraints. 
Also, unlike the case of a scalar resonance, the heavy vector resonance 
is narrow and a simple treatment of the constant width as above is adequate. 

\section{Sensitivity to the Top-SEWS Signal at Future Lepton Colliders}

\subsection{Effective $W$ Approximation}

The effective $W$-boson approximation (EWA)
provides a viable simplification
for high energy processes involving $W$-boson scattering \cite{ewa}.
It is particularly suitable for the calculations under consideration
since the $W$-boson level evaluation for the SEWS signal 
at high energies is adequate. 
The condition for the EWA to be valid is that the invariant
energy scales are all much larger than the $W$ mass, 
namely $s,|t|,|u| \gg M_W$. Besides, it
is more accurate for a longitudinal $W$-boson due to the fact
that the distribution function is largely independent of the
$W$ beam energies. To justify the EWA in our calculation, we
compare our signal evaluation for a heavy scalar resonance
with a full tree-level SM calculation \cite{madg}. 
We define the SM signal for a heavy Higgs boson $\sigma_S$
following the subtraction procedure \cite{baggeretal}
\begin{equation}
\sigma_S(M_H) = \sigma(M_H) - \sigma(M_H=100\ \gev).
\label{sub}
\end{equation}
The interpretation of this procedure is that the signal from
a heavy Higgs boson should be defined as the enhancement over
the perturbative SM expectation with a light ($M_H\sim 100$ GeV) 
Higgs boson, which should be viewed as the irreducible SM
background.

The process of  $e^+ e^- \to Zt\bar t$
followed by $Z\to \nu\bar{\nu}$ leads to the same final state as our
signal. It arises from different physics and needs to be
removed from our consideration.
 We follow the earlier studies
\cite{veto} to introduce a recoil mass against the 
observable $t\bar t$ system defined as
\begin{equation}
M^2_{rec}=s+m^2_{\ttb}-2\sqrt{s}(E_t + E_{\bar t}),
\end{equation}
where $E_t\ (E_{\bar t})$ is the top (anti-top) quark
energy in the $e^+e^-$ CM frame. The recoil-mass
spectrum peaks at $M_Z$ for the background. 
Here and henceforth, we will 
impose $M_{rec} > 200 \gev$ to effectively remove the 
$Z \to \nu\bar{\nu}$ background. We will adopt the same cut for the
$e\nu\ tb$ process as well.
\begin{figure}[tb]
\vskip 0.2in
\centerline{\psfig{file=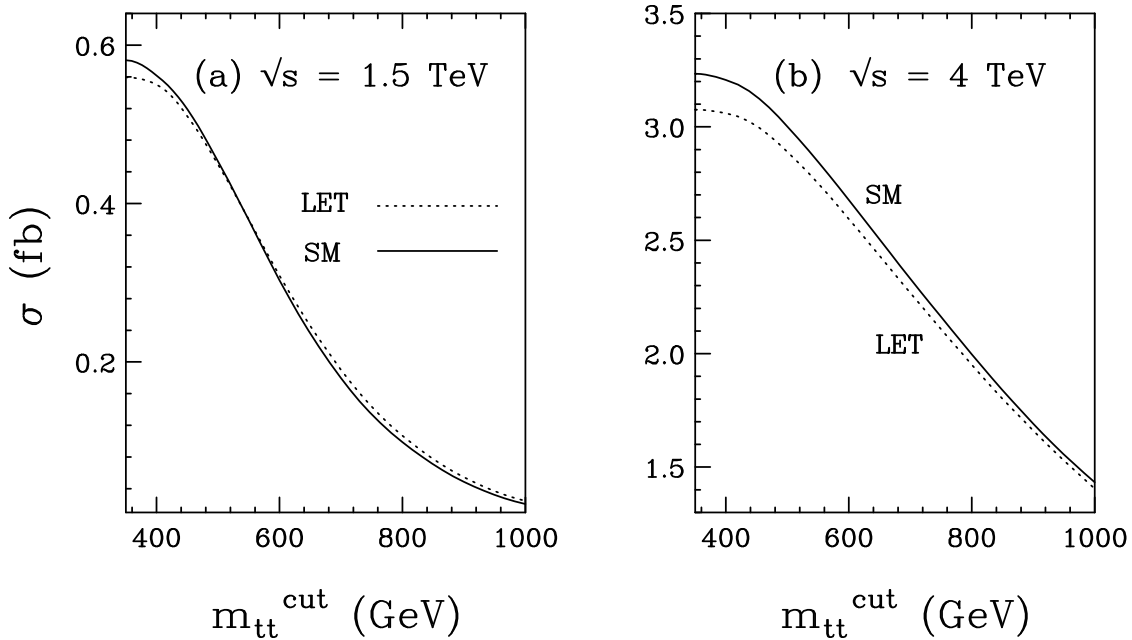,angle=90,width=15cm}}
\caption[]{Production cross sections (in fb) for $e^+e^-\to 
\bar\nu\nu W^+W^- \to \bar\nu\nu \bar t t$ 
versus the invariant mass cut on the top quark pair
for $\mh \to \infty$ at (a) $ \sqrt{s}=1.5 \tev$ and (b) $\sqrt{s}=4 \tev$.
\label{mtt}}
\end{figure}
\begin{figure}[tb]
\vskip 0.2in
\centerline{\psfig{file=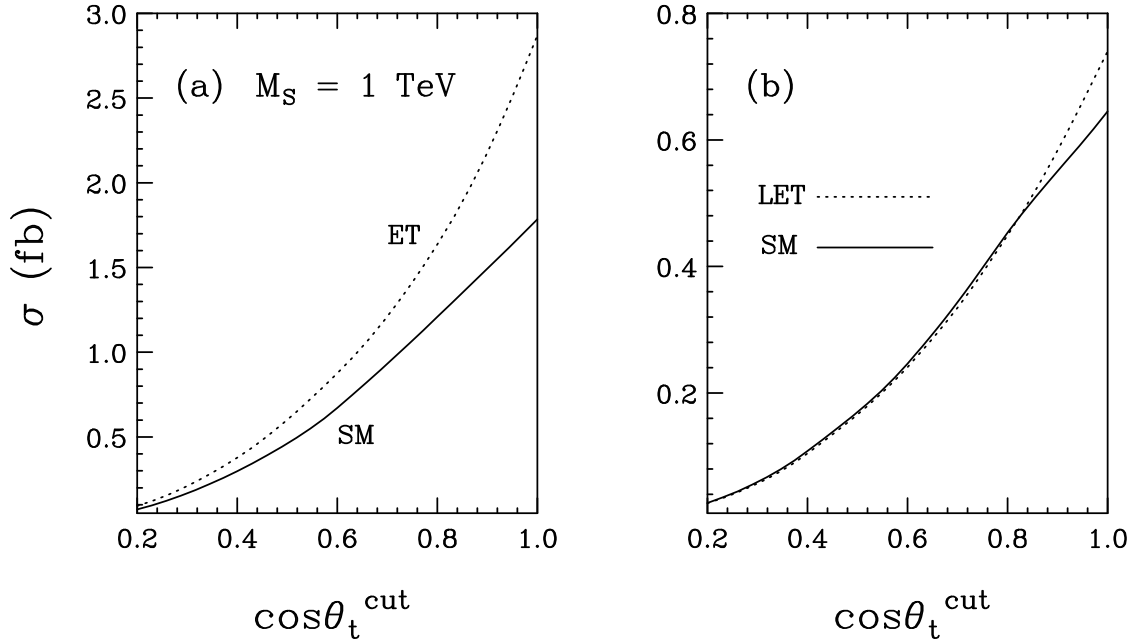,angle=90,width=15cm}}
\caption[]{Production cross sections (in fb) for $e^+e^-\to 
\bar\nu\nu W^+W^- \to \bar\nu\nu \bar t t$ at $\sqrt s=1.5\ \tev$
versus the polar angle cut on the final state $t$ and $\bar t$
(a) for $M_S = 1\ \tev$ and (b) for $\mh \to \infty$.
\label{cos}}
\end{figure}
\begin{figure}[tb]
\vskip 0.2in
\centerline{\psfig{file=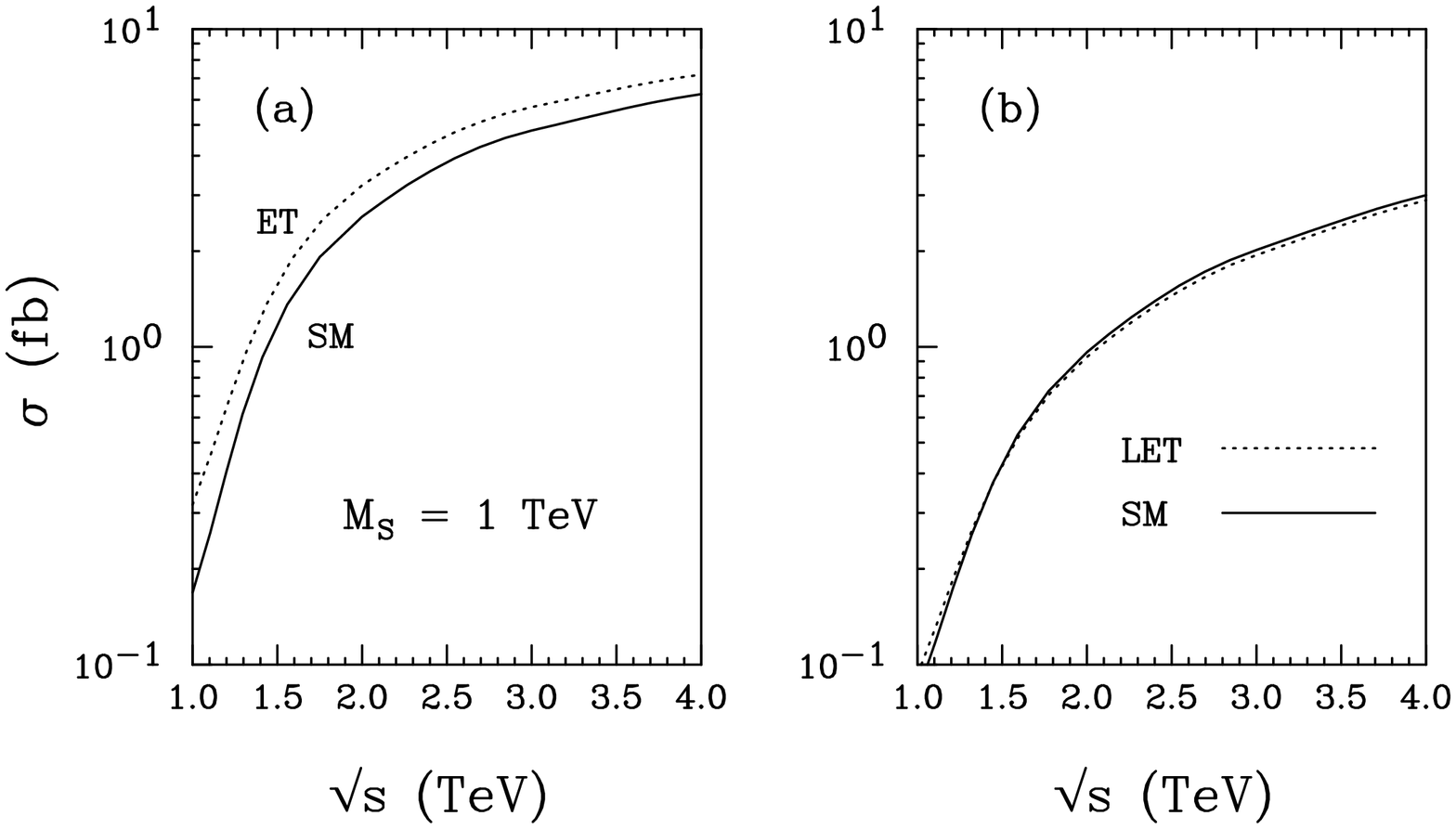,angle=90,width=15cm}}
\caption[]{Production cross sections (in fb) for $e^+e^-\to 
\bar\nu\nu W^+W^- \to \bar\nu\nu \bar t t$ 
versus the CM energy for
(a) $M_S = 1\ \tev$ and (b) $\mh \to \infty$.
\label{sqrts}}
\end{figure}

In Fig.~\ref{mtt}, we present the production cross section for 
$e^+e^-\to \bar\nu\nu W^+W^- \to \bar\nu\nu \bar t t$ 
versus an invariant mass cut ($m_{\ttb}^{cut}$) on the top quark pair
for $\mh \to \infty$ at (a) $e^+e^-$ CM energy $\sqrt s=1.5\ \tev$
and (b) $\sqrt s=4\ \tev$. The solid curves are from the full 
SM calculation based on the subtraction scheme Eq.~(\ref{sub}). 
The dotted curves are the Low Energy Theorem
(LET) result for $w^+w^-\to t\bar t$ by EWA.
The two calculations agree very well, especially for 
higher $m_{\ttb}$ values. 

In Fig.~\ref{cos}, 
we show the cross section for the same process, but
versus the scattering angle cut (${\cos\theta}_t^{cut}$) 
on the final state top quarks. In Fig.~\ref{cos}(a),
we compare the scalar resonance model in Sec.~\ref{scalar}
for $M_S = 1\ \tev$ (dashed curve) with the SM for 
$M_H = 1\ \tev$ (solid curve) based on the subtraction
scheme Eq.~(\ref{sub}). In both cases and henceforth, 
to preserve unitarity near the heavy scalar resonance, 
we adopt the prescription given in Eq.~(\ref{presc}).

Fig.~\ref{cos}(b) compares the case for LET
and the full SM calculation for $\mh \to \infty$. 
We see that the two calculations agree better for
central scattering at low $\cos\theta_t$. These two figures
motivate us to introduce the ``basic acceptance cuts'', 
\begin{equation}
m_{\ttb} > 500\ \gev,\quad |\cos\theta_t|<0.8.
\label{cut1}
\end{equation}
These cuts not only guarantee the good agreement between
the full SM calculation and the EWA that we will adopt
for signal calculations beyond the SM, but are also necessary 
to enhance the signal over the SM backgrounds as we will
discuss later. We impose a similar requirement for the $tb$
process.

Finally we show the comparison
of the two calculations versus the center of mass (CM) 
energy in Fig.~\ref{sqrts} for (a) a $1\ \tev$ scalar
resonance and (b) $\mh \to \infty$, where the basic cuts of 
Eq.~(\ref{cut1}) have been imposed on both $t$ and $\bar t$.
The agreement is generally at order of 50\%
or better for $M_S = 1\ \tev$ and almost perfect for $\mh \to \infty$. 
We consider this to have justified our future use
of the EWA  in the signal calculations beyond
the SM. However, we note that the EWA calculations for
signal processes (\ref{wwtt}) and (\ref{wztb}) do not
properly incorporate the full kinematics of processes
(\ref{eenntt}) and (\ref{eeentb}). We thus simulate the
full kinematics and determine the cut efficiencies using 
the complete SM calculations in the next section.

\subsection{Top-SEWS Signal and SM Backgrounds}

For the SM-like heavy Higgs model adopted in the last section, 
the Higgs boson mass $M_H^{}$ is the only parameter.
We would like to explore more general classes of strongly
interacting electroweak models (SEWS) as outlined in 
Sec.~\ref{sews-top}.
For a generic scalar singlet model, there are three input 
parameters: $M_S^{}$, $g_S^{}$, and $\kappa$ as 
described in section \ref{scalar}. The two couplings can
be traded by the physical partial widths and they are
subject to
the unitarity constraints Eqs.~(\ref{wwgs}, \ref{gsk}, \ref{kappa}).
In terms of the partial widths for $M_S=1\ \tev$, 
these constraints lead to 
\begin{eqnarray}
80\ \gev\lsim&\Gamma_{Sww} &\lsim 600\ \gev,\nonumber\\ 
&\Gamma_{Sww}\Gamma_{S\ttb}&\lsim (300\ \gev)^2,\nonumber\\
&\Gamma_{S\ttb}&\lsim 420\ \gev,
\label{Sbound}
\end{eqnarray}
where we required that partial wave unitarity be satisfied up to 
$\sqrt{s_{ww}}=2\ \tev$.

The three input parameters for the vector model are 
$g_V(=-g_A)$, $\tilde g$, and $a$. 
These can be traded by the physical parameters $\mrho$, 
$\Gamma_{\rho ww}$ (or $\Gamma_{\rho wz}$), 
and $\Gamma_{\rho\ttb}$ (or $\Gamma_{\rho\tbb}$). 
From the constraint of Eq.~(\ref{GVGA}), we get for a neutral 
$\rho^0$ with mass $M_{\rho}=1\ \tev$,
\begin{eqnarray}
\Gamma_{\rho\ttb}\Gamma_{\rho ww}& \lsim& (8\ \gev)^2
\label{tbbound1}
 \end{eqnarray} 
and for a charged $\rho^\pm$ of the same mass, 
\begin{eqnarray}
\Gamma_{\rho tb}\Gamma_{\rho wz}& \lsim& (11\ \gev)^2.
\label{tbbound2}
\end{eqnarray} 

For the $\wl\wl \to \ttb$ signal, the SM backgrounds are
\begin{eqnarray}
e^+e^- &\to& e^+e^-\ t \bar t,
\label{eett}\\
e^+e^- &\to& \bar\nu\nu W^+W^- \to \bar\nu\nu t \bar t.
\label{nntt}
\end{eqnarray}
Process (\ref{eett}) mainly comes from $\gamma\gamma\to \ttb$
and has a large cross section, 
typically about 7.5 fb at $\sqrt s=1.5$ TeV. 
The charged-current process Eq.~(\ref{nntt}) occurs at a lower rate,
about 1.7 fb at $\sqrt s=1.5$ TeV, but it is kinematically
more difficult to separate from the signal. Following 
earlier studies \cite{veto,veto0}, we demand a high transverse
momentum for the final state $t$ and $\bar t$
\begin{equation}
p^{}_T(t) > 150\ {\gev},\quad 
\label{cut2.1}
\end{equation}
which is motivated by the Jacobian peak in the $p_T^{}$ spectrum
from a heavy particle decay. We also require a moderate 
momentum for the $\ttb$ pair
\begin{equation}
30\ \gev < p^{}_T(\ttb) < 300\ {\gev}.
\label{cut2.2}
\end{equation}
This cut forces the final state leptons to be away from
the forward-backward collinear region. Since there are no
energetic electrons in the central region 
for the signal process, we then veto events with final 
state electrons at large angle:
\begin{equation}
E_e  > 50\ {\gev},\quad |\cos\theta_e| < \cos(0.15\ {\rm rad}).
\label{cut3}
\end{equation}
We find that these cuts, along with the basic cuts (\ref{cut1}), are 
very effective to suppress the SM backgrounds. The background
(\ref{eett}) is largely eliminated by the combination of
$p_T^{}(\ttb)$ cut and electron veto. The effects of the cut on the
charged current $t\bar t$ process are summarized in Table~\ref{tt}, 
again using the SM-like heavy Higgs model as the representative 
for the signal. From this table, we find the cut efficiencies 
of Eqs.~(\ref{cut2.1},\ref{cut2.2},\ref{cut3}) for the
signal based on the subtraction scheme (\ref{sub}). They are
\begin{eqnarray} 
\epsilon=83\%\quad {\rm for}\quad  \sqrt{s}=1.5\ \tev;\qquad
\epsilon=84\%\quad {\rm for}\quad  \sqrt{s}=4\ \tev.
\label{efftt}
\end{eqnarray}
For the remainder of this paper we use these cut efficiency figures 
for all the signal calculations within the EWA  in the 
$W_L^{}W_L^{} \to \ttb$ channel.

\begin{table}[tbh]
\begin{center}
\begin{tabular}{l| c| c}\hline\hline
$\sqrt{s}=1.5\tev$ & $\mh=1\tev$ & Background $\mh=0.1\tev$ \\ 
\hline
cuts (\ref{cut1})  &       1.41 fb        &     0.21 fb             \\
cuts (\ref{cut1},\ref{cut2.1},\ref{cut2.2},\ref{cut3}) 
&    1.13 fb            &        0.14 fb    \\ \hline\hline
$\sqrt{s}=4\tev$   & $\mh=1\tev$ & Background $\mh=0.1\tev$ \\ 
\hline
cuts (\ref{cut1})  &       8.86 fb         &     0.98 fb            \\
cuts (\ref{cut1},\ref{cut2.1},\ref{cut2.2},\ref{cut3})
&    7.20 fb         &        0.55 fb    \\ \hline\hline
\end{tabular}
\end{center}
\vspace{0.1in}
\caption[]{Cross sections of process (\ref{nntt})
for a SM-like heavy Higgs (1 TeV)
and a light Higgs (0.1 TeV) with different cuts. This gives
the cut efficiencies for Top-SEWS signals as in 
Eq.~(\ref{efftt}).}
\label{tt}
\end{table}

\begin{figure}[tb]
\vskip 0.2in
\centerline{\psfig{file=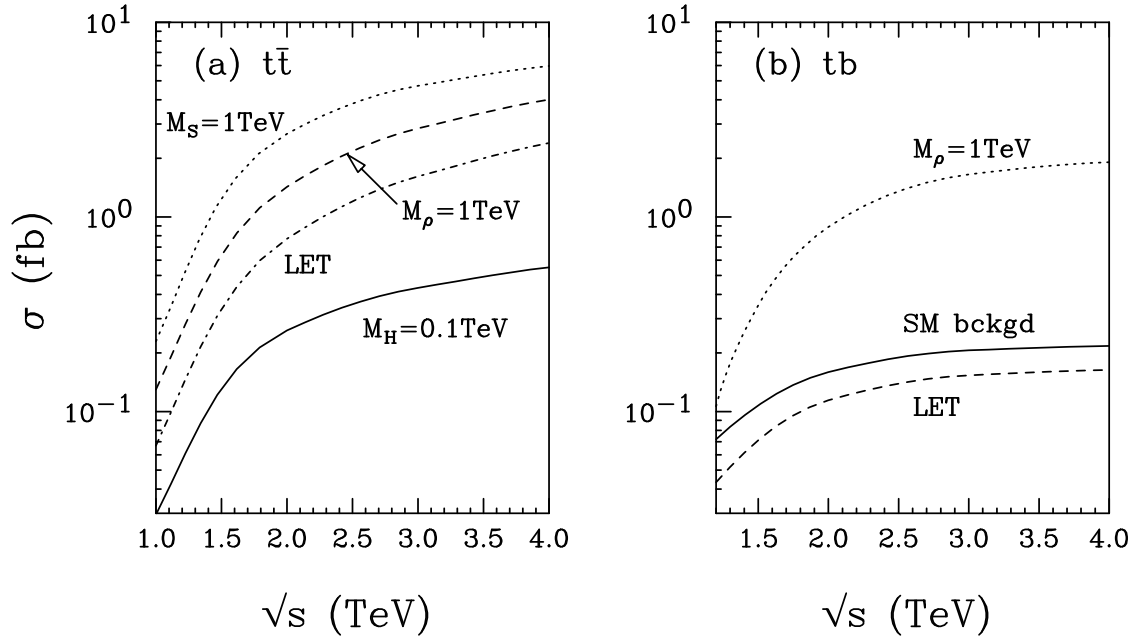,angle=90,width=15cm}}
\caption[]{Production cross sections (in fb) for (a) $e^+e^-\to 
\bar\nu\nu W^+W^- \to \bar\nu\nu \bar t t$ versus $\sqrt s$
for $\mh = 0.1\ \tev$ (solid), LET (dot-dash),  
 scalar singlet model (dots), and vector model (dashes), 
and for (b) $e^+e^-\to e \nu\ WZ \to e \nu\  tb$ 
versus $\sqrt s$ for the vector model (dots), LET (dashes)
and SM background (solid) with the full cuts imposed. 
We have used the
model parameters as in Eqs.~(\ref{paratt}) and (\ref{paratb}).
\label{tot}}
\end{figure}

For the process $\wl\zl \to tb$, the irreducible background
is from
\begin{eqnarray}
e^+e^- \to e \nu\ WZ \to e \nu\ tb.
\label{entb}
\end{eqnarray}
Again, the largest contribution comes from the photon induced
process $e\gamma \to \nu\ t\bar b$, where the other electron goes
along the beam direction after emitting a collinear photon. 
For this case, we impose the cuts (\ref{cut1}) and
(\ref{cut2.1}). Since there is a final state electron in the signal
process with a typical transverse momentum of $M_Z/2$,
we require an electron tagging at a finite angle
\begin{equation}
|\cos\theta_e| < \cos(0.15\ {\rm rad}).
\label{cut4}
\end{equation}
This electron tagging is very effective to remove the SM
background, and more so at higher energies.
The combination of the above cuts significantly reduces the
$e\gamma$ process and brings the background to a level below 
the signal rate.

To determine the signal efficiency of the electron tagging, 
we calculate the process
\begin{eqnarray}
e^+ e^- \to \bar \nu e^- \rho^+.
\end{eqnarray}
We find
\begin{eqnarray} 
\epsilon=93\%\quad {\rm for}\quad  \sqrt{s}=1.5\ \tev;\qquad
\epsilon=61\%\quad {\rm for}\quad  \sqrt{s}=4\ \tev,
\end{eqnarray}
which will be used for other signal calculations  in the  
$W_L^{}W_L^{} \to tb$ channel based on the EWA. 
The lower tagging efficiency at 4 TeV is due to the fact
that the final state electron becomes more forward at
higher energies.

%
%

We now present the production cross sections for the signal and 
background versus the CM energy $\sqrt s$ after
the cuts. Fig.~5(a) is for 
$e^+e^-\to \bar\nu\nu W^+W^- \to \bar\nu\nu\ \bar t t$ process with the 
full cuts (\ref{cut1}, \ref{cut2.1}, \ref{cut2.2}, \ref{cut3}) 
for the scalar model (dots), vector model (dashes), 
LET (dot-dash), and 
the irreducible background $\mh = 0.1\ \tev$ (solid).
For illustration, we have taken the model parameters to be
\begin{eqnarray}
M_S=1\ \tev,\ \Gamma_{Sww}=493\ \gev,\ \Gamma_{S\ttb}=50\ \gev,
\nonumber\\  
\mrho=1\ \tev,\ \Gamma_{\rho ww} = 50\ \gev,\ \Gamma_{\rho \ttb} = 1.3\ \gev. 
\label{paratt}
\end{eqnarray}
Figure~\ref{tot}(b) shows the production cross sections for 
$e^+e^- \to e \nu\ WZ \to e \nu\ tb$, 
with the cuts (\ref{cut1}, \ref{cut2.1}, \ref{cut4}) for
the vector model (dots), LET (dashes), and the
SM background (solid). The model parameters are taken as
\begin{eqnarray}
\mrho=1\ \tev,\ \Gamma_{\rho ww} = 50\ \gev,\ 
\Gamma_{\rho tb} = 2.5\ \gev.
\label{paratb}
\end{eqnarray} 
We see from Fig.~\ref{tot} that our cuts effectively 
suppress the SM background below the signal for both channels. 
The signal rate at a 1.5 TeV linear collider is typically
between 0.5 and 1.2 fb.

\subsection{Results and discussion}
\begin{figure}[tb]
\vskip 0.2in
\centerline{\psfig{file=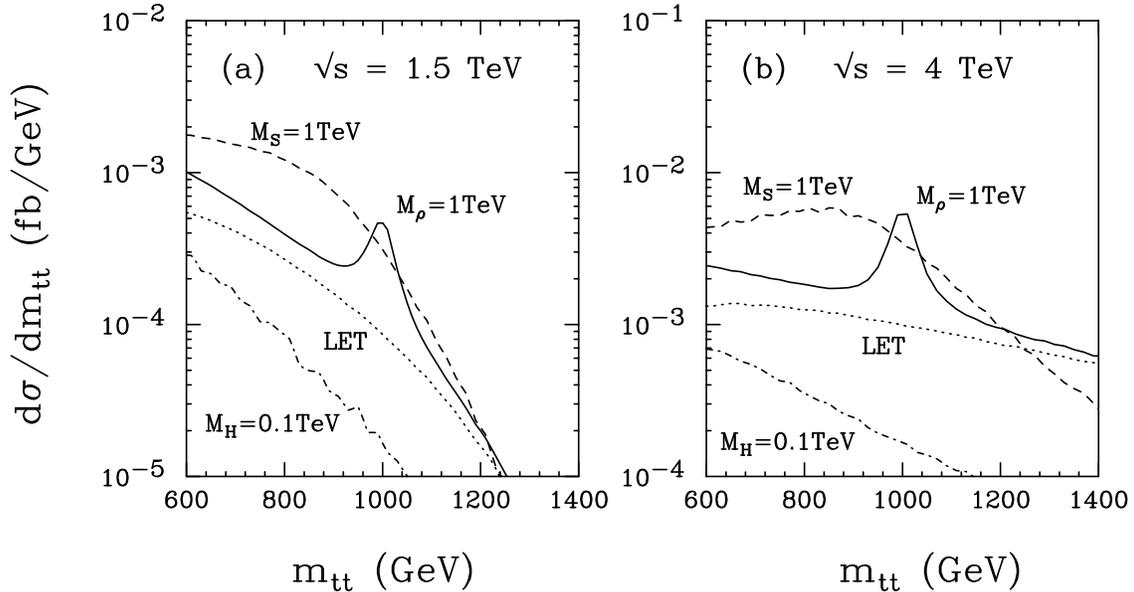,angle=90,width=15cm}}
\caption[]{Signal and background differential cross sections 
versus $\ttb$ invariant mass for (a) 
$\sqrt{s}= 1.5\ \tev$ and (b)
  $\sqrt{s}= 4\ \tev$: SM background by $\mh = 0.1\ \tev$ 
(dot-dash), LET (dots),  scalar model (dashes), 
and vector model (solid). We have used the model parameters
as in Eq.~(\ref{paratt}).
All cuts are imposed and the
branching fraction for $t\to bjj'$ is included.    
\label{inv1}}
\end{figure}
\begin{figure}[tb]
\vskip 0.2in
\centerline{\psfig{file=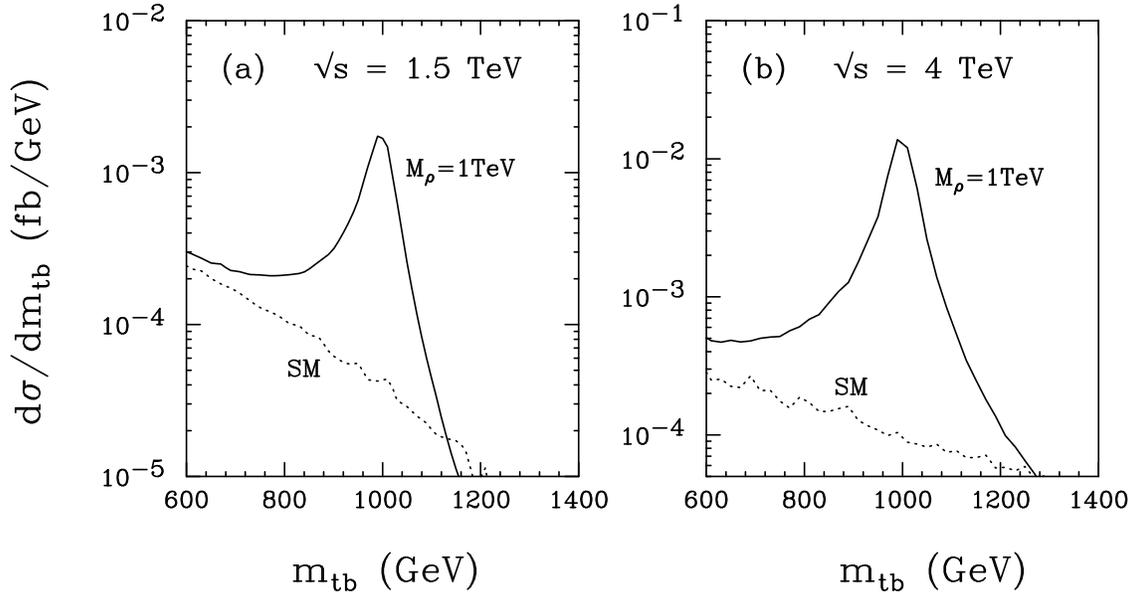,angle=90,width=15cm}}
\caption[]{Signal and background cross sections (in fb) versus $tb$
  invariant mass ($m_{tb}$) for (a) $\sqrt{s}= 1.5\ \tev$ and (b)
  $\sqrt{s}= 4\ \tev$: SM (dots) and vector model (solid). 
We have used the model parameters as in Eq.~(\ref{paratb}). 
All cuts are imposed and the
branching fraction for $t\to bjj'$ is included.      
\label{inv2}}
\end{figure}
\begin{figure}[tb]
\vskip 0.2in
\centerline{\psfig{file=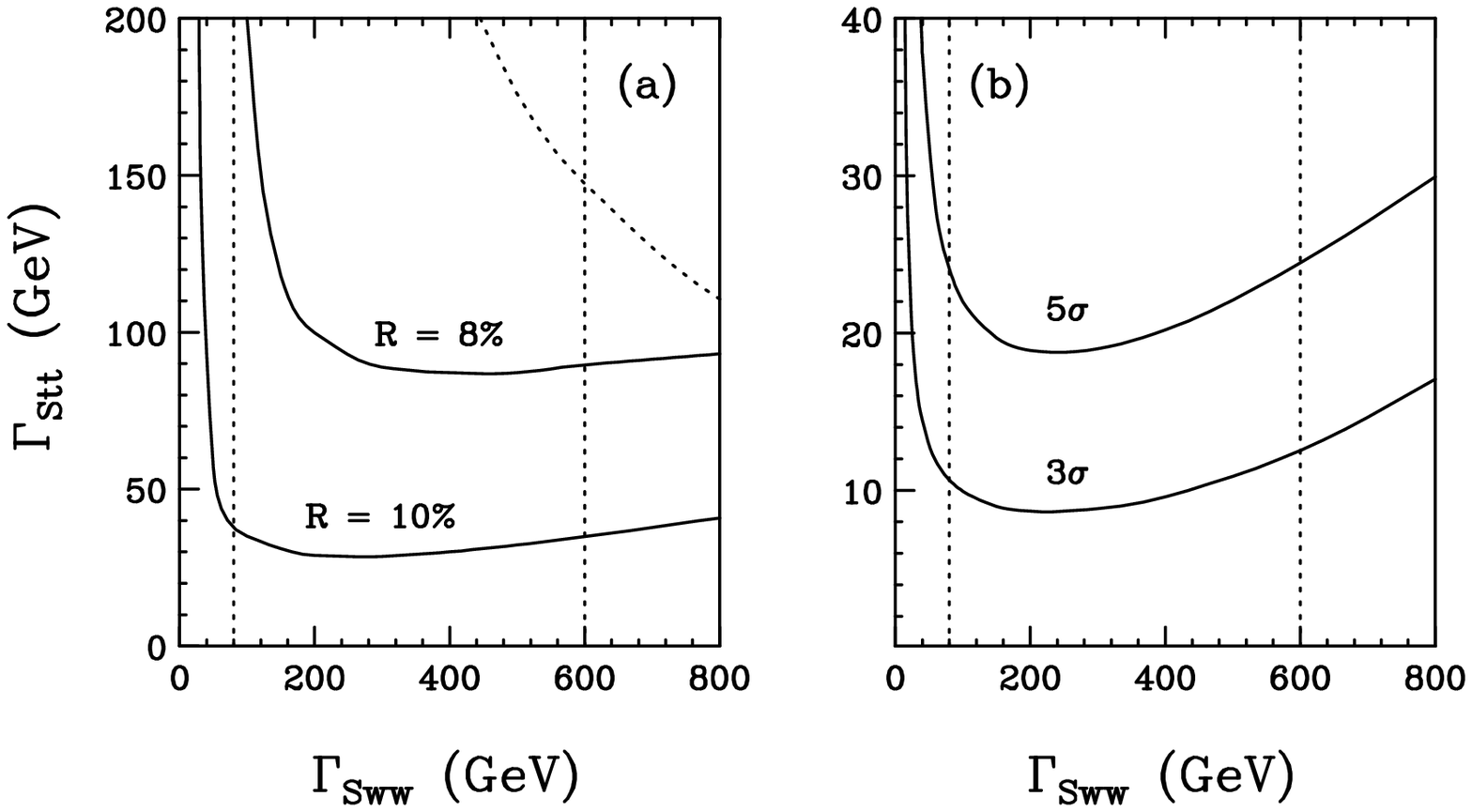,width=15cm}}
\caption[]{At a linear collider with
$\sqrt s=1.5\ \tev$ and a luminosity of 200 $\fbi$,
(a) accuracy contours of determination of the partial widths
for the scalar resonance model with $M_S=1$ TeV;
(b) statistical significance contours to distinguish
a 1 TeV scalar from LET. Dotted lines are unitarity constraints given
in Eq.~(\ref{Sbound}).
\label{sens}}
\end{figure}

For the remainder of our analysis, we consider the top quark to decay
hadronically $t\to bW \to b jj'$ with a branching fraction 
approximately $70\%$. In doing so, we assume that
we can fully reconstruct the $t$ and $\bar t$ events, and
that there are no isolated 
central electrons from the decay to confuse 
our electron vetoing and tagging requirements. 
We also assume a data sample of 200 $\fbi$ at $\sqrt s=1.5$ TeV. 
We can see from Fig.~\ref{tot} that we may reach a clear
signal observation for both a scalar and a vector
resonance in the $t\bar t$ channel. Although the LET
channel has relatively lower rate, it may nevertheless lead to 
about $35$ events after including the branching fraction,
which is well above the background expectation.
For the $tb$ channel, both signal and background
rates are slightly smaller than the $t\bar t$ case,
but the vector signal rate is still sufficiently 
above background to be observable. The LET amplitude
in this case is below background
as was already seen in the model discussions.

It is important to know how well one can reconstruct the
signal and contrast it with the background beyond simple event counting. 
In Fig.~\ref{inv1} we show the distributions for the
signal and background versus $\ttb$
  invariant mass ($m_{\ttb}$) for (a) $\sqrt{s}= 1.5\ \tev$ and (b)
  $\sqrt{s}= 4\ \tev$ with the SM background represented by 
$\mh = 0.1\ \tev$ (dot-dash), LET signal (dots),  
 scalar model (dashes), and vector model (solid). The
model parameters are taken as in Eq.~(\ref{paratt}).
All cuts are imposed and
the branching fraction for $t\bar t$ decay is included.
We see a broad resonance around $M_S=1\ \tev$ in scalar model due to 
its large width. For the vector case, however, the width is 
narrower in general and the signal 
peaks above the scalar model for $M_{\rho}=1\ \tev$ in a 
distinctive manner. The LET model has relatively low rate and 
has no particular structure.
In Fig.~\ref{inv2} we show the same signal and background 
distributions, but for the $tb$ channel versus the
invariant mass ($m_{tb}$). The model parameters are
taken as in Eq.~(\ref{paratb}).
Clearly, the observation of a signal in Fig.~\ref{inv2} 
would provide decisive information to discriminate a scalar
resonance from a vector. The authors of Ref.~\cite{peskin}  
use, instead, the top-quark polarization information 
to distinguish a scalar from a vector resonance.

We next consider the extent to which one can probe the couplings for
different models. Fig.~\ref{sens}(a) shows the accuracy contours
to determine the partial decay widths (thus the
couplings) in the scalar resonance model with 
 $M_S=1$ TeV at $\sqrt s=1.5\ \tev$ and a luminosity of 200 $\fbi$. 
The statistical accuracy 
for the cross section measurement is defined as 
\begin{eqnarray}
R&=&{\sqrt {S+B} \over S}
\nonumber
\end{eqnarray}
where $S$ and $B$ are the number of signal and background events 
respectively and we have ignored the experimental systematic 
effects. We find that partial widths can be probed 
up to about $10\%$ accuracy for $\Gamma_{S \ttb}\approx 50\ \gev$ 
within the unitarity bounds shown as dotted 
lines (Eq.~\ref{Sbound}). 
In the region above the curve labeled $R=8\%$, the partial widths can be 
determined to $8\%$ or better accuracy. 

Also shown in 
Fig.~\ref{sens}(b) are the statistical significance contours
to distinguish a 1 TeV scalar signal from the LET amplitudes.
Here we define an appropriate standard deviation as
\begin{eqnarray}
\sigma&=&\{ \sigma(\Gamma_{S\ttb},\Gamma_{Sww})-\sigma(0,0)\}{\cal L} 
\over {\sqrt {\sigma(0,0) {\cal L}}}
\nonumber
\end{eqnarray}
where $\sigma(\Gamma_{S\ttb},\Gamma_{Sww})$ is the cross section
in the scalar model for the given widths, $\sigma(0,0)$ is 
the non-resonance cross section with $g_S=\kappa=0$, 
and $\cal L$ is the luminosity. 
It is easy to see that the scalar model is distinguishable at 
the 5$\sigma$ level
from the non-resonant case within the unitarity limits 
given in Eq.~(\ref{Sbound}).

\begin{figure}[tb]
\vskip 0.2in
\centerline{\psfig{file=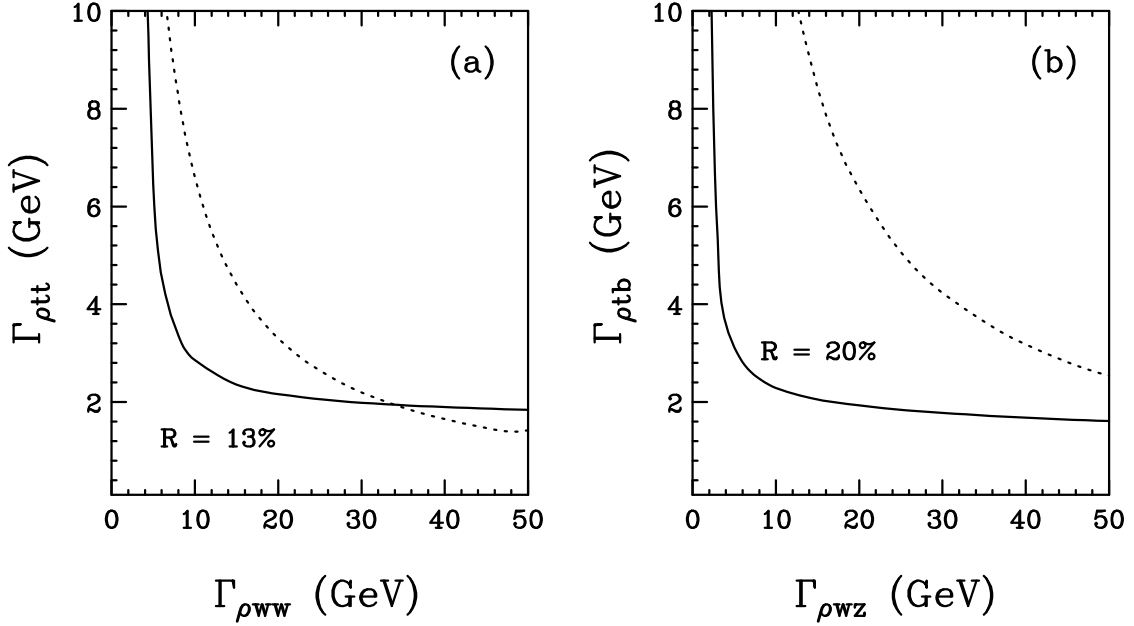,width=15cm}}
\caption[]{At a linear collider with
$\sqrt s=1.5\ \tev$ and a luminosity of 200 $\fbi$,
determination of the partial widths
for the vector resonance model with $M_\rho=1$ TeV
(a) $13\%$ accuracy contour for $W^+W^- \to \rho^0 \to \ttb$
channel; (b)  $20\%$ accuracy contour 
for $W^{\pm}Z \to \rho^{\pm} \to \tbb(\bar t b)$.
Dotted lines are from the current low energy constraints 
in Eqs.~(\ref{tbbound1}) and (\ref{tbbound2}).
\label{senrho}}
\end{figure}
We show the same analysis for the vector model in Fig.~\ref{senrho},
where we present the sensitivity to the vector resonance model with 
$M_\rho=1$ TeV at $\sqrt s=1.5\ \tev$ and a luminosity of 200 $\fbi$,
for (a) $13\%$ accuracy determination of the partial
widths for  $W^+W^- \to \rho^0 \to \ttb$ and (b)  $20\%$ accuracy
determination of the partial widths for 
$W^{\pm}Z \to \rho^{\pm} \to \tbb\ (\bar t b)$.
We found in  Fig.~\ref{senrho}(a) that a $13\%$ accuracy determination 
of the partial widths is possible for the neutral vector process 
in the region allowed by the current bounds given
in Eq.~(\ref{tbbound1}). 
For the charged process we have a smaller 
signal rate, nevertheless, we found in Fig.~\ref{senrho}(b) that 
we can still determine the partial widths to 
$ 20\%$ accuracy in the region allowed by current constraints, 
Eq.~(\ref{tbbound2}). 

\section{Conclusion}
We have studied the couplings of the top quark to TeV resonances that 
occur in models in which the electroweak symmetry is broken by a strong 
interaction. Our study is motivated by the fact that in many models 
the top quark plays a special role in the breaking of electroweak 
symmetry. 

We have explored
the possibility that the new physics is realized as a scalar 
resonance or a vector resonance near 1 TeV, as well as 
the case where the resonance is beyond reach (LET). 
For the scalar model, we used partial wave unitarity 
to constrain the parameters as shown in Sec.~\ref{scalar}.  
For the vector model, we used current experimental data to constrain 
the parameter space as presented in Sec.~\ref{vector}. We found 
within these constraints, that signal rates in these models at future 
lepton colliders can be well above the SM backgrounds after judicial 
cuts are applied. The models can be distinguished from each other by 
systematically studying two different final states. Typically 
one would see a broad scalar resonance $S$, or a relatively narrow 
vector resonance $\rho^0$ in the $t\bar t$ channel 
and a narrow vector resonance $\rho^\pm$ in the $tb$ channel.
In particular, we illustrated our results at a 1.5 TeV linear
collider with an integrated luminosity of 200 fb$^{-1}$.
One expects to observe about 120 events for either $S\to t\bar t$ 
or $\rho^{0,\pm}\to t\bar t,\ tb$ including final state 
identification. 
Even if the lower-lying resonances are inaccessible
at the collider, one can still observe a statistically
significant deviation as predicted by the LET
for the $t\bar t$ final state
from the perturbative SM expectation.
Finally, the leading partial decay widths, which 
characterize the coupling strengths, can be statistically 
determined to about $10\%$ level.

{\it Acknowledgments}: 
The work of T.H. and Y.J.K. was supported in part by the US DOE under 
contract No. DE-FG02-95ER40896 
and in part by the Wisconsin Alumni Research Foundation. 
The work of A.L. was supported in part by the RFBR grants 
99-02-16558 and 00-15-96645.
The work of G.V. was supported in part by DOE under contact number 
DE-FG02-92ER40730. G.V. thanks the Special Research Centre for the 
Subatomic Structure of Matter at the University of Adelaide for their 
hospitality and partial support while part of this work was completed.

\newpage

\appendix
\section{SM amplitudes using the Equivalence\\ Theorem}
\label{app}
In this appendix, we present the standard model amplitudes obtained using 
the Equivalence Theorem for the processes $w^-w^+ \to \ttb$, $zz \to \ttb$, 
and $zw^+ \to \tbb$. From these amplitudes one can derive the 
(LET) amplitudes by taking the large Higgs mass limit, ignoring the 
electroweak couplings, and setting $m_b=0$.

\vspace{10mm}
{\underline {$w^-w^+ \to \ttb$}} 
\begin{eqnarray}
{\cal M^{\pm\pm}}_{H}(ww\to \ttb)&=&
\pm{ {\mh}^2 \over  (s-{\mh}^2) } {m_t\sqrt s \over v^2}\beta_t
\nonumber \\
{\cal M^{\pm\mp}}_{H}(ww\to \ttb)&=&0
\\
{\cal M^{\pm\pm}}_b(ww\to \ttb)&=&
 \mp{2m_t^3 \over v^2 \sqrt s}
{\beta_t+\cos\theta \over 1+\beta_t \cos\theta-2m_t^2/s}
\nonumber \\
{\cal M^{\pm\mp}}_{b}(ww\to \ttb)&=& {m_t^2 \over v^2}{(1\pm\beta_t)
\sin\theta\over 1+\beta_t\cos\theta -2 m_t^2/s}
\\
{\cal M^{\pm\pm}}_{\gamma}(ww\to \ttb)&=& 
\pm{2 g^2 x_W Q}{ m_t \over \sqrt s}\cos\theta
\nonumber \\
{\cal M^{\pm\mp}}_{\gamma}(ww\to \ttb)&=& 
\mp g^2 x_W Q \sin\theta
\\
{\cal M^{\pm\pm}}_{Z}(ww\to \ttb)&=&
\pm 2 g_V g^2_Z({1 \over 2}-x_W){ m_t \over \sqrt{s}} \cos\theta
\nonumber\\
{\cal M^{\pm\mp}}_{Z}(ww\to \ttb)&=& (-g_V\mp g_A\beta_t)g^2_Z({1 \over 2}-x_W)
\sin\theta
\end{eqnarray}
where 
\begin{displaymath}
\beta_t = (1-4{m_t^2/ s})^{1 \over 2}, \quad
x_W=\sin^2\theta_W, \quad 
g_Z={g\over \cos\theta_W},\quad 
g_V= {1 \over 2}T_3 - Q x_W, \quad {\rm and} \quad 
g_A= - {1 \over 2}T_3.
\end{displaymath}

Taking the limit $M^2_H \gg s$ and ignoring the terms of order $g^2$,
we find to leading order,
\begin{eqnarray}
{\rm LET:}\quad {\cal M^{\pm\pm}}_{LET}(ww\to \ttb)&=&
\mp {m_t \sqrt s\over v^2}\beta_t
\nonumber \\
{\cal M^{\pm\mp}}_{LET}(ww\to \ttb)&=& {m_t^2 \over v^2}{(1\pm\beta_t)
\sin\theta\over 1+\beta_t\cos\theta -2 m_t^2/s}.
\nonumber 
\end{eqnarray}

\vspace{10mm}
{\underline {$zz \to \ttb$}}
\begin{eqnarray}
{\cal M^{\pm\pm}}_{H}(zz\to \ttb)&=&
\pm{ {\mh}^2 \over  (s-{\mh}^2) } {m_t\sqrt s \over v^2}\beta_t
\nonumber \\
{\cal M^{\pm\mp}}_{H}(zz\to \ttb)&=& 0
\\
{\cal M^{\pm\mp}}_{top}(zz\to \ttb)&=&
 \pm {2m_t^3 \over v^2 \sqrt s}
{\beta_t+\cos\theta \over 1+\beta_t \cos\theta}
\nonumber \\
{\cal M^{\pm\mp}}_{top}(zz\to \ttb)&=&
- {m_t^2 \over v^2}{\sin\theta \over 1+\beta_t\cos\theta}. 
\end{eqnarray}
\begin{eqnarray}
{\rm LET:}\quad{\cal M^{\pm\pm}}_{LET}(zz\to \ttb)&=&
 \mp {m_t \sqrt s\over v^2}\beta_t
\nonumber \\
{\cal M^{\pm\mp}}_{LET}(zz\to \ttb)&=&
 -{m_t^2\over v^2}
{\sin\theta\over 1+\beta_t\cos\theta}.
\end{eqnarray}
Again we keep only the leading order terms and 
ignore the top-quark exchange diagram for 
${\cal M^{\pm\pm}}_{LET}(zz\to~\ttb)$.

\vspace{10mm}
{\underline {$zw^+ \to \tbb$}}
\begin{eqnarray}
{\cal M^{++}}_{W^+}(zw^+\to \tbb)&=&
-{\sqrt{2} \over 4}g^2{ m_t  \over \sqrt{s}}\beta_m \cos\theta
\nonumber \\
{\cal M^{--}}_{W^+}(zw^+\to \tbb)&=&{\cal M^{+-}}_{W^+}(zw^+\to \tbb)=0
\nonumber \\
{\cal M^{-+}}_{W^+}(zw^+\to \tbb)&=&
{\sqrt{2} \over 4}g^2\beta_m \sin\theta
\end{eqnarray}
where $\beta_m=(1-m_t^2/s)^{1/2}$.
\begin{eqnarray}
{{\cal M}^{++}}_{top}(z w^+ \to \tbb)&=&
{\sqrt 2 m_t^3\over v^2 \sqrt{s}} { \beta_m (1+\cos\theta)
\over [\beta_m^2 (1-\cos\theta)+2m_t^2/s]}  \nonumber \\
{\cal M}^{--}_{top}(z w^+ \to \tbb)&=&{\cal M}^{+-}_{top}(z w^+ \to\tbb)=0
\nonumber \\
{{\cal M}^{-+}}_{top}(z w^+ \to \tbb)&=&
-{\sqrt 2 m_t^2 \over v^2}{\beta_m  \sin\theta
\over {[\beta_m^2 (1-\cos\theta)+2m_t^2/s]}}  
\end{eqnarray}
where ${\cal M}_{b}(z w^+ \to \tbb)$ is proportional to $m_b$ and is neglected.

\begin{eqnarray}
{\rm LET:}\quad{{\cal M}^{++}}_{LET}(z w^+\to \tbb)&=&
{\sqrt 2 m_t^3\over v^2 \sqrt{s}} { \beta_m (1+\cos\theta)
\over [\beta_m^2 (1-\cos\theta)+2m_t^2/s]}  \nonumber \\
{{\cal M}^{-+}}_{LET}(z w^+\to \tbb)&=&
-{\sqrt 2 m_t^2 \over v^2}{\beta_m  \sin\theta
\over {[\beta_m^2 (1-\cos\theta)+2m_t^2/s]}}  .
\end{eqnarray}


\newpage

\begin{thebibliography}{99}

\bibitem{sews} For a review, see {\it e. g.}, 
M. Chanowitz, Ann. Rev. Nucl. Part. Sci. {\bf 38}, 323 (1988).

\bibitem{topcolor} C. Hill, Phys. Lett. {\bf B266}, 419 (1991);
{\it ibid.} {\bf B345}, 483 (1995);
E. Eichten and K. Lane, Phys. Lett. {\bf B352}, 382 (1995).

\bibitem{topss} B. Dobrescu and C. Hill, 
Phys. Rev. Lett. {\bf 81}, 2634 (1998); 
R. S. Chivukula, B. Dobrescu, H. Georgi, and C. Hill,
Phys. Rev. {\bf D59}, 075003 (1999); 
G. Burdman and N. Evans, Phys. Rev. {\bf D59}, 115005 (1999).

\bibitem{liz} E. H. Simmons, hep-ph/9908488. 

\bibitem{eqt} B. W. Lee, C. Quigg, and H. B. Thacker, Phys. Rev. {\bf D16},
1519 (1977); M.S. Chanowitz and M.K. Gaillard, Nucl. Phys. {\bf B261},
379 (1985); Y.P. Yao and C.P. Yuan, Phys. Rev. {\bf D38}, 2237 (1988); 
J. Bagger and C. Schmidt, Phys. Rev. {\bf D41}, 264 (1990); H.
Veltman, Phys. Rev. {\bf D41}, 2294 (1990); H.-J. He, Y.-P. Kuang, and
X. Li, Phys. Rev. Lett. {\bf 69}, 2619 (1992); H.-J. He and
W.B. Kilgore, Phys. Rev. {\bf D55}, 1515 (1997).

\bibitem{cfh}M. Chanowitz, M. Furman and I. Hinchliffe, 
Nucl. Phys. {\bf B153}, 402 (1979).

\bibitem{ewa} M. Chanowitz and M.K. Gaillard, Phys. Lett. {\bf B142},
  85 (1984); G. Kane, W. Repko and W. Rolick, Phys. Lett. {\bf B148}, 
367 (1984); S. Dawson, Nucl. Phys. {\bf B249}, 42 (1985).


\bibitem{russel} R.P. Kauffman, Phys. Rev. {\bf D41}, 3343 (1990).

\bibitem{steveg} M. Gintner and S. Godfrey, in the Proceedings of
{\it New Directions for High-Energy Physics}, June 25-July 12, 1996,
Snowmass, p.~824.

\bibitem{eff} F. Larios, E. Malkawi and C.P. Yuan,
Acta Phys. Polon. {\bf B27}, 3741 (1996); F. Larios and C.-P. Yuan, 
Phys. Rev. {\bf D55}, 7218 (1997);
Jose Wudka, talk given at 
{\it the 5th International Conference on Physics Beyond the 
Standard Model}, Balholm, Norway, 29 Apr - 4 May 1997, 
hep-ph/9706434.

\bibitem{timb} T. Barklow, in the Proceedings of
{\it New Directions for High-Energy Physics}, June 25-July 12, 1996, 
Snowmass, p.~819; T. Han, hep-ph/9910495.

\bibitem{cp2} F. Larios, Tim Tait, and C.-P. Yuan, 
Phys. Rev. {\bf D57}, 3106 (1998).

\bibitem{muon}V. Barger, M. Berger, J. Gunion and T. Han,
talk given at {\it ITP Conference on Future High-energy Colliders}, 
Santa Barbara, CA, 21-25 Oct 1996, hep-ph/9704290.

\bibitem{mikep} T. Barklow \etal, in the Proceedings of
{\it New Directions for High-Energy Physics}, June 25-July 12, 1996, 
Snowmass, p.~735.

\bibitem{peskin} E. R. Morales and M. E. Peskin, hep-ph/9909383.

\bibitem{let} M.S. Chanowitz, M. Golden and H. Georgi, Phys. Rev. Lett.
{\bf 57}, 2344 (1986); Phys. Rev. {\bf D35}, 1149 (1987). 

\bibitem{order2} A. Dobado and M. Herrero, Phys. Lett. {\bf B228},495 
(1989); {\it ibid.} {\bf B233}, 505 (1989); J. Donoghue and C. Ramirez, 
Phys. Lett. {\bf B234}, 361 (1990); 
S. Dawson and G. Valencia, Nucl. Phys. {\bf B348}, 
23 (1991); {\it ibid.} {\bf B352}, 27 (1991).

\bibitem{xinmin} R. D. Peccei, S. Peris, and X. Zhang, 
Nucl. Phys. {\bf B349}, 305 (1991).

\bibitem{ac} T. Appelquist and M. Chanowitz, Phys. Rev. Lett. 
{\bf 59}, 2405 (1987).

\bibitem{mvw}
W.~Marciano, G.~Valencia and S.~Willenbrock, Phys. Rev. {\bf D40}, 1725 (1989).

\bibitem{bdv}
J.~Bagger, S.~Dawson and G.~Valencia, Nucl. Phys. {\bf B399}, 364 (1993).

\bibitem{baggeretal} J. Bagger, V. Barger, K. Cheung, J. Gunion,
T. Han, G. Ladinsky, R. Rosenfeld, C.-P. Yuan, Phys. Rev.
{\bf D49}, 1246 (1994); {\it ibid.} {\bf D52}, 3878 (1995).

\bibitem{seymour} M.H. Seymour, Phys. Lett. {\bf B354}, 409 (1995).

\bibitem{scottres} G.~Valencia and S.~Willenbrock,
Phys. Rev.  {\bf D42}, 853 (1990); 
{\it ibid.} {\bf D46}, 2247 (1992).

\bibitem{scott} S. Jager and S. Willenbrock, 
Phys. Lett. {\bf B435}, 139 (1998);
R. S. Chivukula, Phys. Lett. {\bf B439}, 389 (1998).

\bibitem{bagsm} J.~Bagger, T.~Han and R.~Rosenfeld,
FERMILAB-CONF-90-253-T, published in {\it Snowmass Summer Study 1990},
p. 208.

\bibitem{bess} R.~Casalbuoni \etal, Phys. Lett. {\bf B155}, 
95 (1985); Nucl. Phys. {\bf B282}, 235 (1987); {\it ibid.} {\bf B310}, 
181 (1988); Phys. Lett. {\bf B249}, 130 (1990); {\it ibid.} {\bf B253},
275 (1991).


\bibitem{fy} K.~Fujikawa and A.~Yamada, Phys. Rev. {\bf D49}, 5890 (1994).

\bibitem{lpy}F.~Larios, M.~A.~Perez and C.~P.~Yuan,
Phys. Lett. {\bf B457}, 334 (1999).

\bibitem{dv} S.~Dawson and G.~Valencia, Phys. Rev. {\bf D53}, 1721 (1996).

\bibitem{pdb} C.~Caso {\it et al.},
Eur. Phys. J. {\bf C3}, 1 (1998).

\bibitem{hv} A.~A.~El-Hady and G.~Valencia, Phys. Lett. {\bf B414}, 173 (1997).

\bibitem{yuan} T.~Tait and C.~P.~Yuan, hep-ph/9710372.

\bibitem{bessre}D.~Dominici,
 Riv. Nuovo Cim. {\bf 20}, 1 (1997) [hep-ph/9711385].

 \bibitem{madg} For the full SM calculations,
we have made use of the Madgraph package by
T. Stelzer and W.F. Long, 
 Comput. Phys. Commun. {\bf 81}, 357 (1994).

\bibitem{veto} V. Barger, K. Cheung, T. Han and R.J.N. Phillips,
Phys. Rev. {\bf D52}, 3815 (1995);
V. Barger, M. Berger, J. Gunion and T. Han,
Phys. Rev. {\bf D55}, 142 (1997).

\bibitem{veto0} V. Barger, K. Cheung, T. Han and R.J.N. Phillips,
Phys. Rev. {\bf D42}, 3052 (1990).

\end{thebibliography}
\end{document}